\documentclass[12pt,aps,prb,preprint]{revtex4}   

\usepackage{amsmath}    
\usepackage{graphicx}   
\usepackage{subfigure}
\begin{document}

\title{Observation of magnetic field lines in the vicinity of a superconductor with the naked eye 
}
\author{Yoshihiko Saito}
\affiliation{Osaka Science Museum, Osaka-shi 530-0005 Japan}
\email{saito@sci-museum.jp}   
\date{\today}

\begin{abstract}
Meissner effect and pinning effect are clearly observed with the naked eye. A GdBaCuO high-temperature superconductor (HTS) disk fabricated by Nippon Steel Corporation, a 100mm cubic NdFeB sintered magnet, and iron wires coated by colored are used. When the HTS is put in the magnetic field of the magnet, it can be observed by the wires that the magnetic field lines are excluded from the superconductor (Meissner effect) as well as are pinned in the superconductor@(pinning effect). 

\end{abstract}

\maketitle

\section{Introduction}
\begin{figure}
  \centering
    \includegraphics[width=60mm]{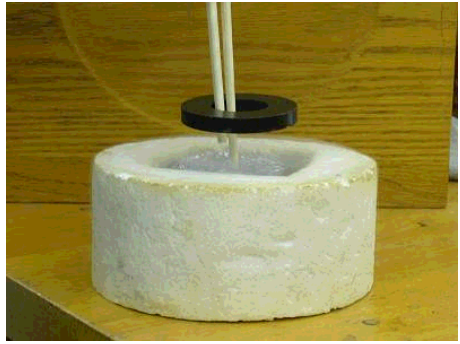}
    \caption{ Magnet floating above a superconductor. 
A HTS is cooled in liquid nitrogen. A NdFeB sintered ring magnet is repelled by the HTS.
The horizontal movement of the magnet is restricted by a pair of chopsticks. }
    \label{fig:mag_sc1}
\end{figure}

\begin{figure}
     \subfigure[]{\includegraphics[height=40mm]{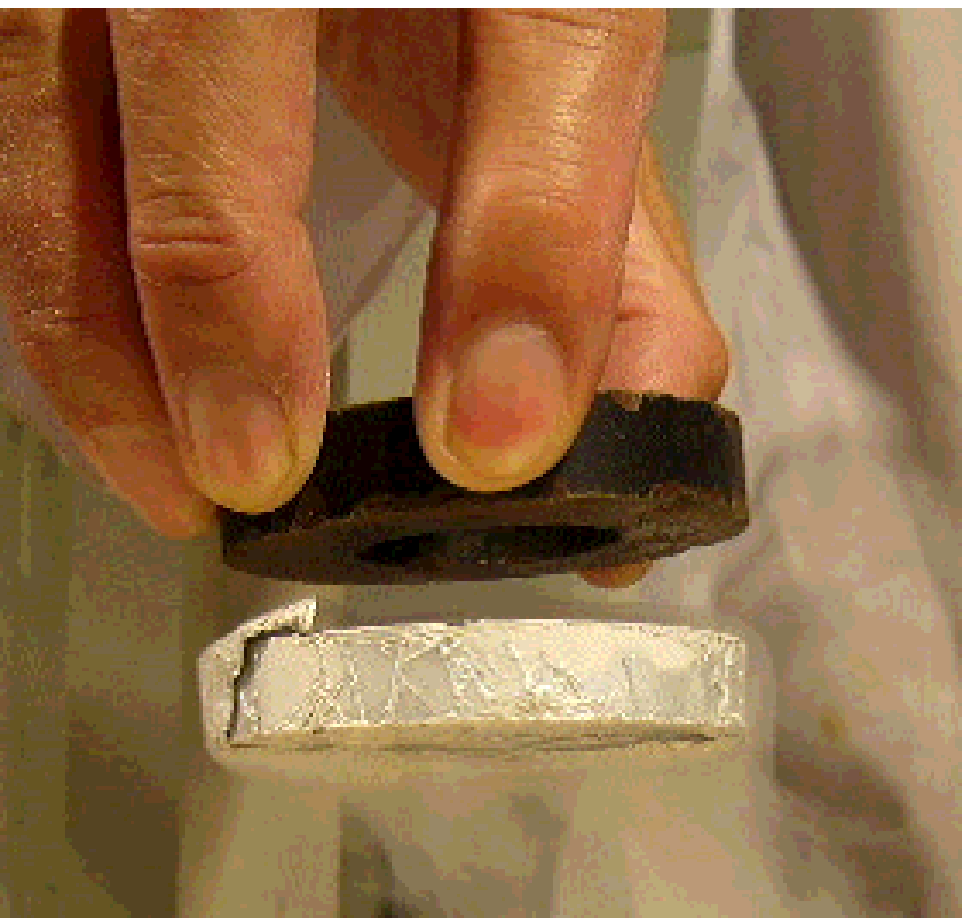}\label{fig:m_s_2b}}
     \subfigure[]{\includegraphics[height=40mm]{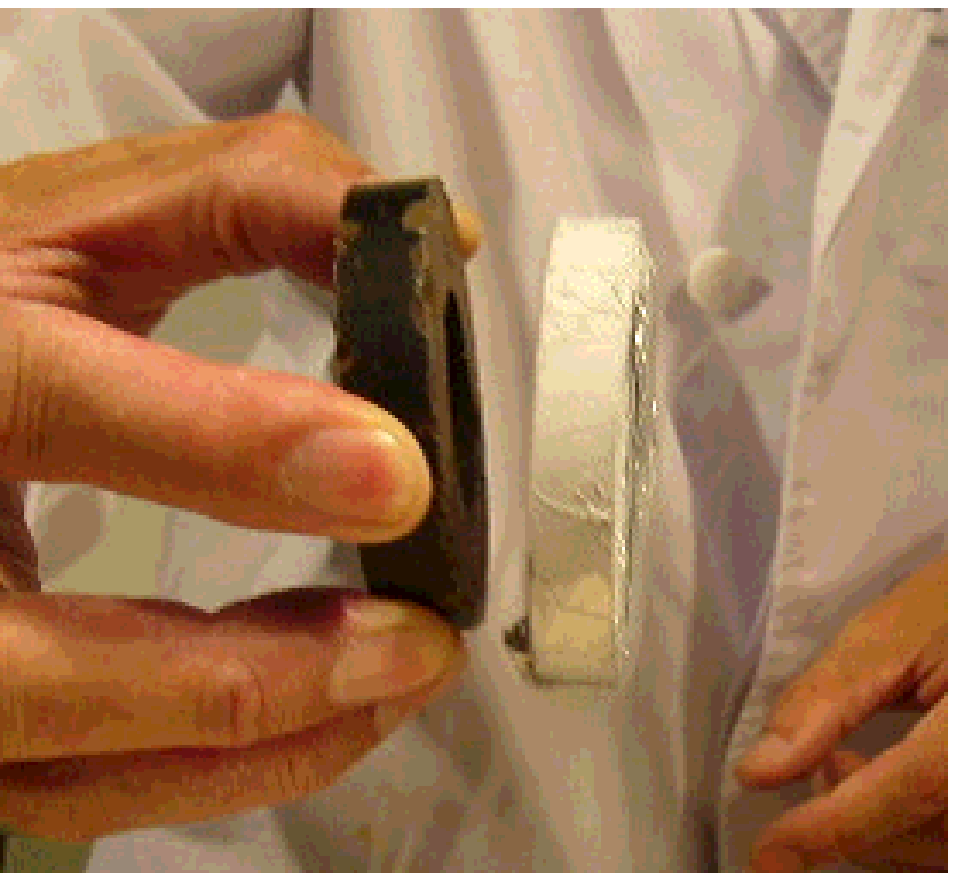}\label{fig:m_s_2c}}
    \subfigure[]{\includegraphics[height=40mm]{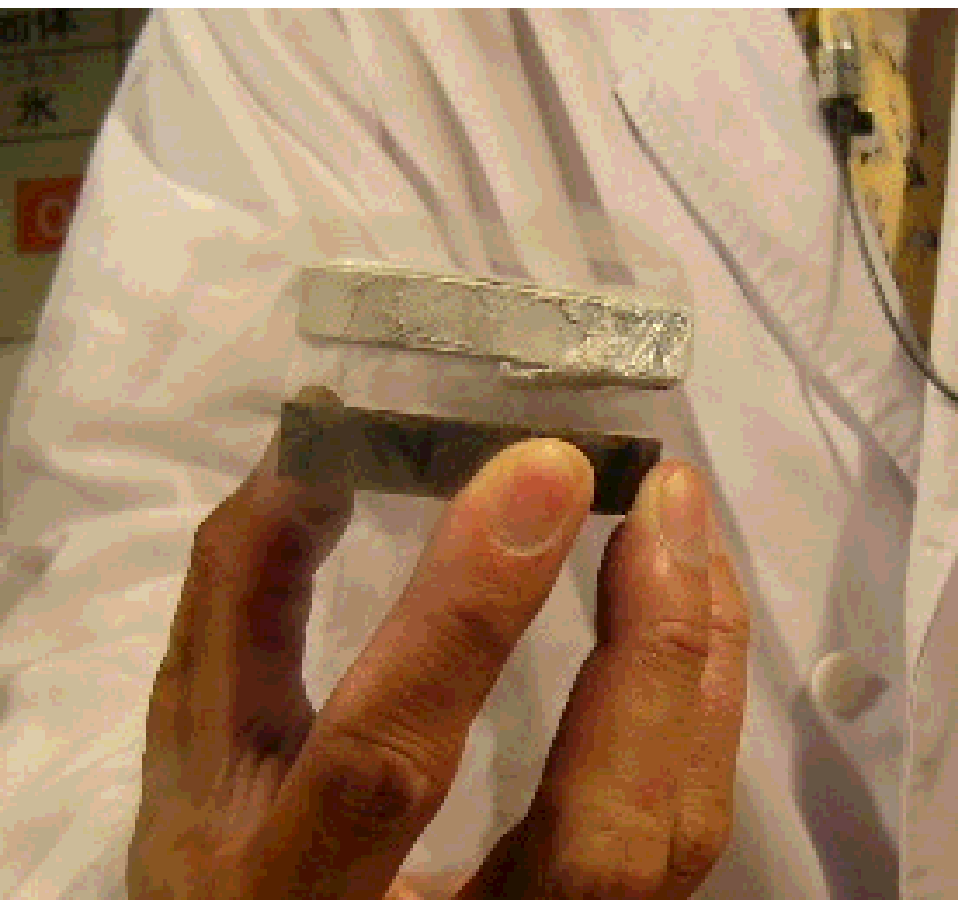}\label{fig:m_s_2d}}
   \caption{
Superconductor bound by a magnet at a fixed distance. (a) A HTS is suspended by a NdFeB sintered ring magnet (the held black object). (b) (c) The magnet is capsized. The relative position of the HTS and the magnet is kept at a fixed distance.}
    \label{fig:mag_sc2}
\end{figure}

 A demonstration of superconductivity with a high-temperature superconductor (HTS) was demonstrated  as early as the year of the HTS discovery \cite{demo1}. Soon, apparatuses for the demonstration were developed\cite{demo2}$^,$\cite{demo3}, and rigid levitation and suspension of the HTSs by magnets was discussed taking account of pinning effect\cite{demo4}. Today, even beginners can enjoy large-scale phenomena of superconductivity and have strong impression. For example, Fig.~\ref{fig:mag_sc1} and Fig.~\ref{fig:mag_sc2} are the demonstrations with a GdBaCuO HTS disk fabricated by Nippon Steel Corporation\cite{sc}. The disk size is diameter 60mm and height 10mm. In Fig.~\ref{fig:mag_sc1}, a NdFeB sintered ring magnet is repelled by the HTS and the horizontal movements of the magnet are restricted by a pair of chopsticks, so the magnet is floating above the HTS. Fig.~\ref{fig:mag_sc2} is a continuous shot, namely (a) suspended HTS, and (b) (c) capsized magnet with the HTS. It can be seen that the relative position of the HTS and the magnet is kept at a fixed distance. If these two phenomena are shown successively, the demonstration will be more magical and more mysterious, because they are interactions of the same HTS and the same magnet. However, no one could see the essence of superconductivity.
 
 The phenomenon that a magnet floats above a HTS (Fig.~\ref{fig:mag_sc1}) is caused by Meissner effect\cite{Meissner}. The Meissner effect is the phenomenon that magnetic field lines are excluded from superconductors. In the HTS, the magnetic field of the magnet is canceled by induced supercurrents and mirror images of each pole are produced; therefore, the magnet is repelled by the HTS. The phenomenon that a HTS is bound to a magnet at a fixed distance (Fig.~\ref{fig:mag_sc2}) is caused by pinning effect and the Meissner effect\cite{demo4}$^,$\cite{demo7.0}$^,$\cite{demo7.1}. The pinning effect is the phenomenon that a vortex lattice is pinned by the defects in the superconductor crystal. A vortex lattice is formed by the interactions between the quantized magnetic flux tubes penetrating the superconductor with normal state core\cite{tonomura}$^,$\cite{Vortex}. The flux tube is called vortex. At the macroscopic level, the pinned effect is the phenomenon that the magnetic field lines cannot move and change inside a superconductor. In Fig.~\ref{fig:mag_sc2}, while the Meissner effect acts repulsively, the interaction between the magnetic field lines pinned in the superconductor and those of the magnet acts attractively, therefore, the superconductor and the magnet do not get too close as well as too far each other.

If the demonstration is scientific one but not an astonishment show, the behavior of the magnetic field lines should be mentioned and it is desirable that the magnetic field lines are visualized.

Though quantitative considerations had been made\cite{demo4}$^,$\cite{demo7.0}$^,$\cite{demo7.1}, no one seems to know a demonstration such that the beginners observe the magnetic field lines in the phenomena. There are explanations of the phenomena with some illustrations or animations, but unfortunately it may be difficult for the beginners to accept them, because they are unrealistic. Therefore, it is desirable that the beginners observe the magnetic field lines with the naked eye.

In 1989, A. Tonomura et al. visualized the magnetic field lines in superconductivity at the vortex level by the use of electron holography and digital phase analysis\cite{tonomura}. This is a historic observation for basic research as well as for industrial applications. For the beginners, however, this method is too complicated to use or understand.   

The author developed a very simple observation of large-scale magnetic field lines with the naked eye, and visualized the Meissner effect\cite{saito}. For the case of the pinning effect, he did not observe the magnetic field lines, which are pinned in a HTS. The observation of the pinning effect was not complete. In this article, observations of the magnetic field lines for the Meissner effect as well as for the pinning effect will be shown.

\section{Visualization of magnetism in superconductivity}
\begin{figure}[h]
  \centering
    \includegraphics[width=60mm]{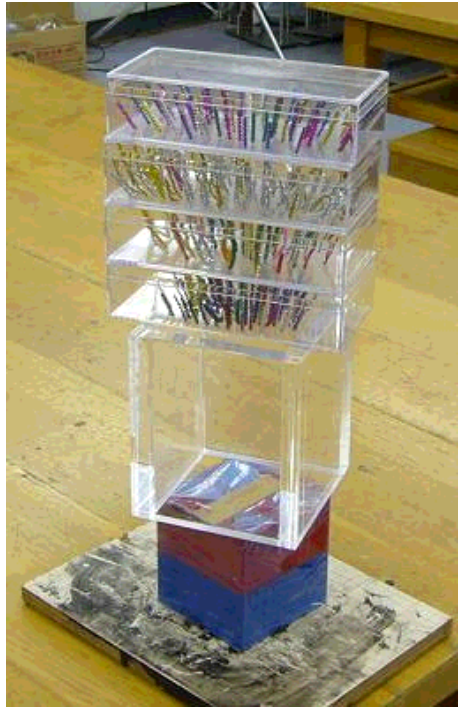}
    \caption{Magnetic field lines diverging from a 100mm cubic NdFeB sintered magnet of magnetic flux density about 0.5T at the surface.
The magnetic field lines are shown by the iron wires in the acrylic cases. }
    \label{fig:magnetic_field}
\end{figure}

In this section, we will observe the magnetic field lines in the vicinity of the HTS in the state of Fig.~\ref{fig:mag_sc1}, Fig.~\ref{fig:mag_sc2}, and so on. The HTS will be put in the magnetic field which is caused by a 100mm cubic NdFeB sintered magnet of magnetic flux density about 0.5T at the surface (Fig.~\ref{fig:magnetic_field}). The magnetic field is shown by iron wires which are coated by colored vinyl to observe more clearly. The wires are almost free to move and rotate under the influences of the magnetic field and gravity. The direction any wire points is tangent to the curve of a magnetic field line and the density of the wires corresponds to that of the magnetic field lines, the magnetic flux. The principle of this method is the same as that of the well-known observation with iron filings. The effect of gravity on the wires is negligible for qualitative understandings unless we observe too far from the magnet. They are enough to supply qualitative information. The magnetic field lines for the Meissner effect as well as for the pinning effect are observed as follows.

\newpage

\subsection{ The Meissner effect}
\begin{figure}
  \centering
    \subfigure[]{\includegraphics[width=90mm]{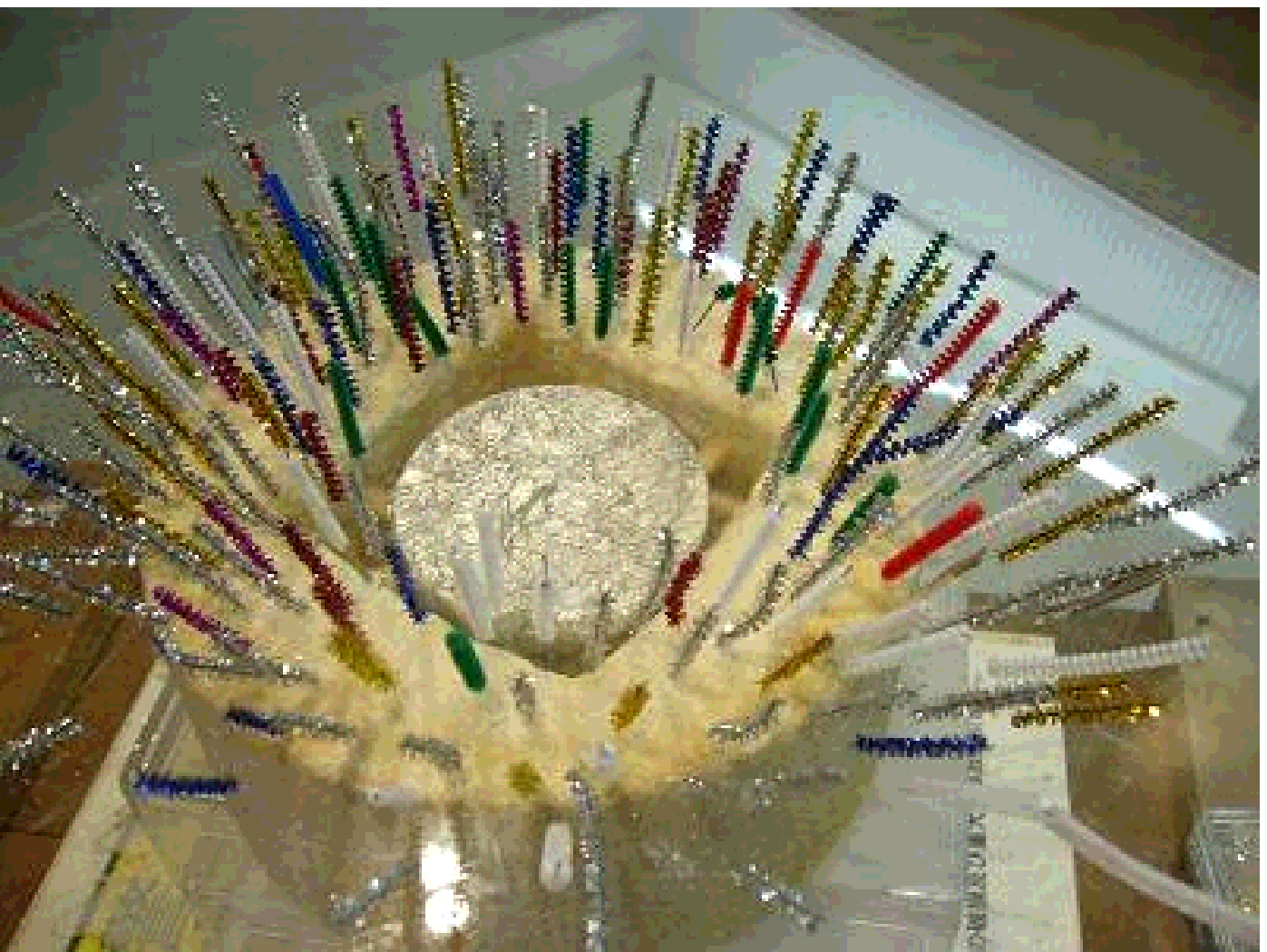}\label{fig:meissner_1}}
    \subfigure[]{\includegraphics[width=60mm]{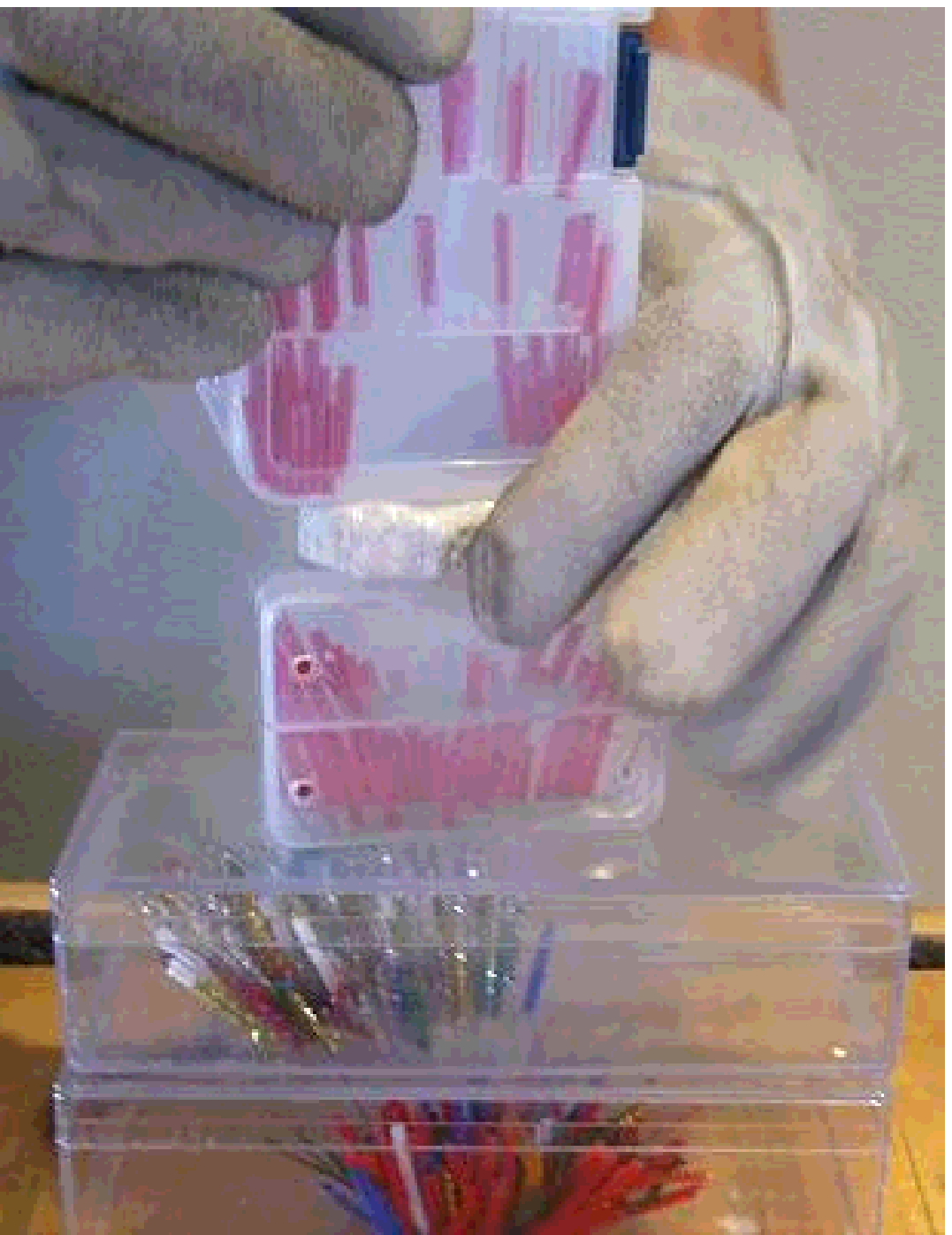}\label{fig:meissner_2}}
	\caption{Meissner effect shown by wires. The wires are: (a) on a transparent plastic board on the top of the HTS;
(b) in the transparent plastic cases divided by multistage on the top and the bottom of the HTS. There exists the 100mm cubic NdFeB sintered magnet below the HTS and the HTS is in the external magnetic field caused by the magnet. There is no wire on the HTS, while the magnetic field lines in the vicinity of the HTS are shown by the wires. It can be seen that the external magnetic field liens are excluded from the HTS. } 
    \label{fig:meissner}
  \centering
    \subfigure[]{\includegraphics[width=60mm]{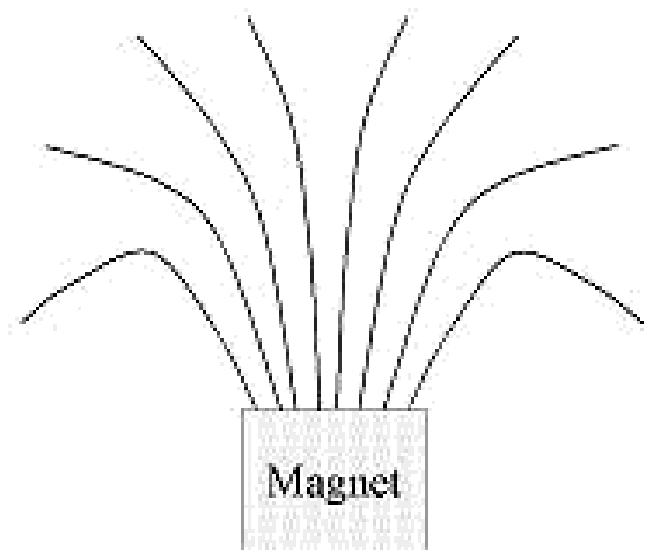}\label{fig:mfl_1}}
    \subfigure[]{\includegraphics[width=60mm]{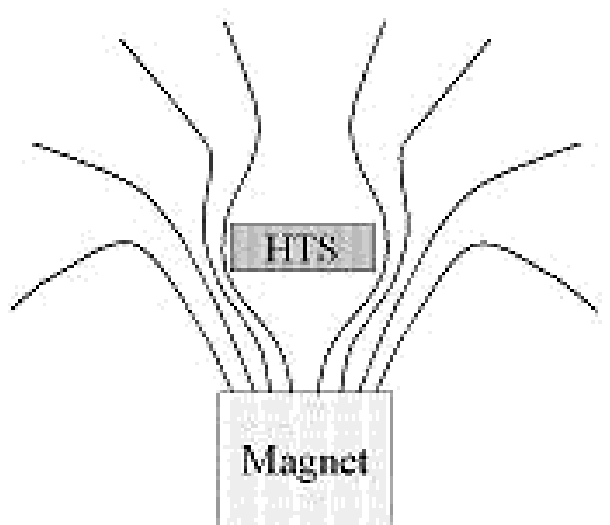}\label{fig:mfl_2}}
	\caption{Diagram of the Meissner effect. (a) the free magnetic field lines diverging from a magnet. If the HTS in Fig.~\ref{fig:mag_sc1} is put in the magnetic field, (b) the magnetic field lines are excluded from a superconductor, namely the Meissner effect.}  
    \label{fig:sc_meissner}
\end{figure}

Fig.~\ref{fig:meissner} shows the magnetic field lines in the vicinity of the HTS in the state of Fig.~\ref{fig:mag_sc1}. The external magnetic field is caused by the 100mm cubic NdFeB sintered magnet under the HTS and the external magnetic field lines are excluded from the HTS like Fig.~\ref{fig:sc_meissner}. In Fig.~\ref{fig:meissner_1}, the wires are on a transparent plastic board on the top of the HTS. It is shown by the wires that the magnetic field lines do not exist on the HTS. In Fig.~\ref{fig:meissner_2}, the wires are in the transparent plastic cases divided by multistage on the top and the bottom of the HTS. It is shown by the wires that the magnetic field lines do not penetrate the HTS. It can be seen that the external magnetic field lines do not exist in the HTS like Fig.~\ref{fig:sc_meissner}. This is the very Meissner effect.

\newpage 

The cooling method of the HTS is the following. The HTS is put in near zero external magnetic field while the HTS is cooled in liquid nitrogen, i.e., at the transition from the normal state to the superconducting state, so that Meissner effect will occur in the HTS. This cooling method is important as mentioned at the next subsection.

\subsection{The pinning effect}
\begin{figure}
    \begin{center}
    \subfigure[]{\includegraphics[height=40mm]{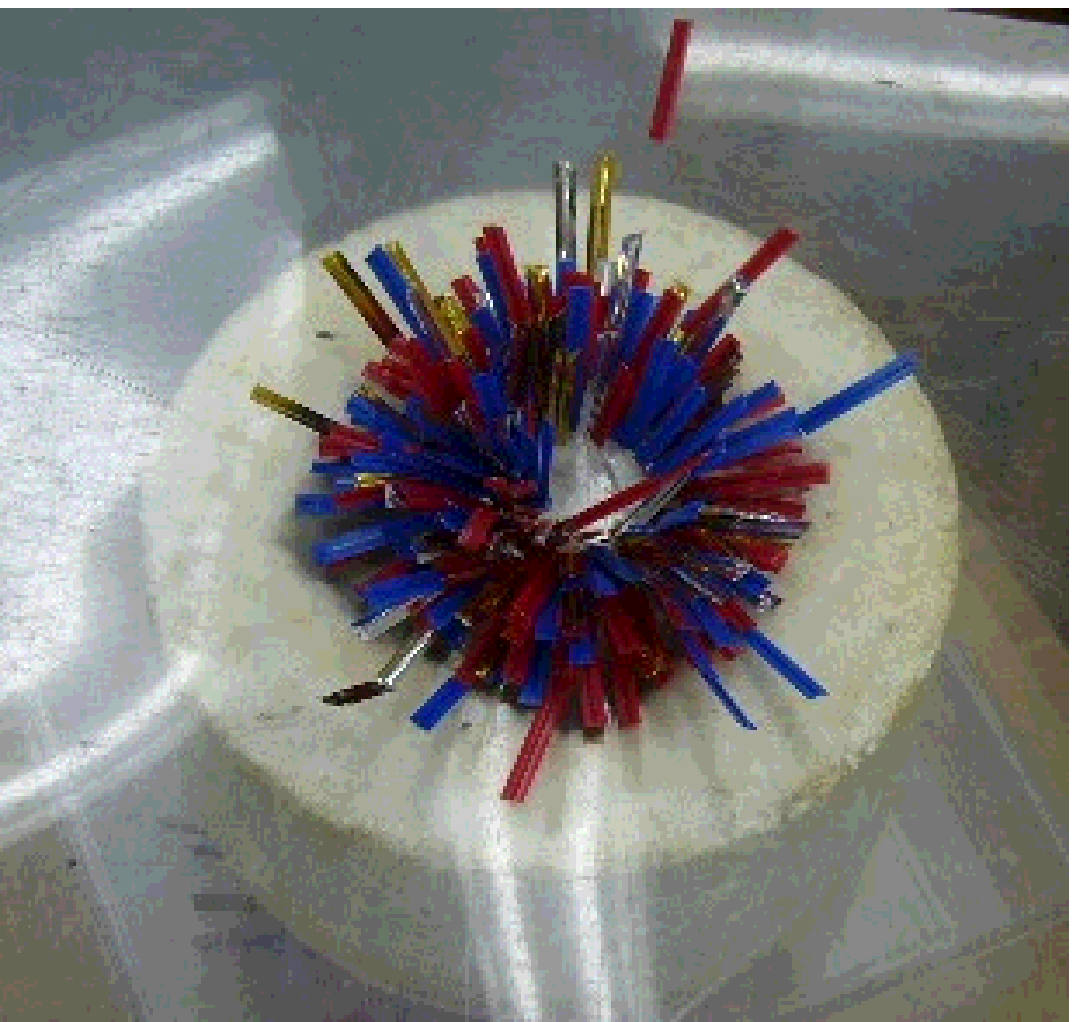}\label{fig:pinning_1a}}
    \subfigure[]{\includegraphics[height=40mm]{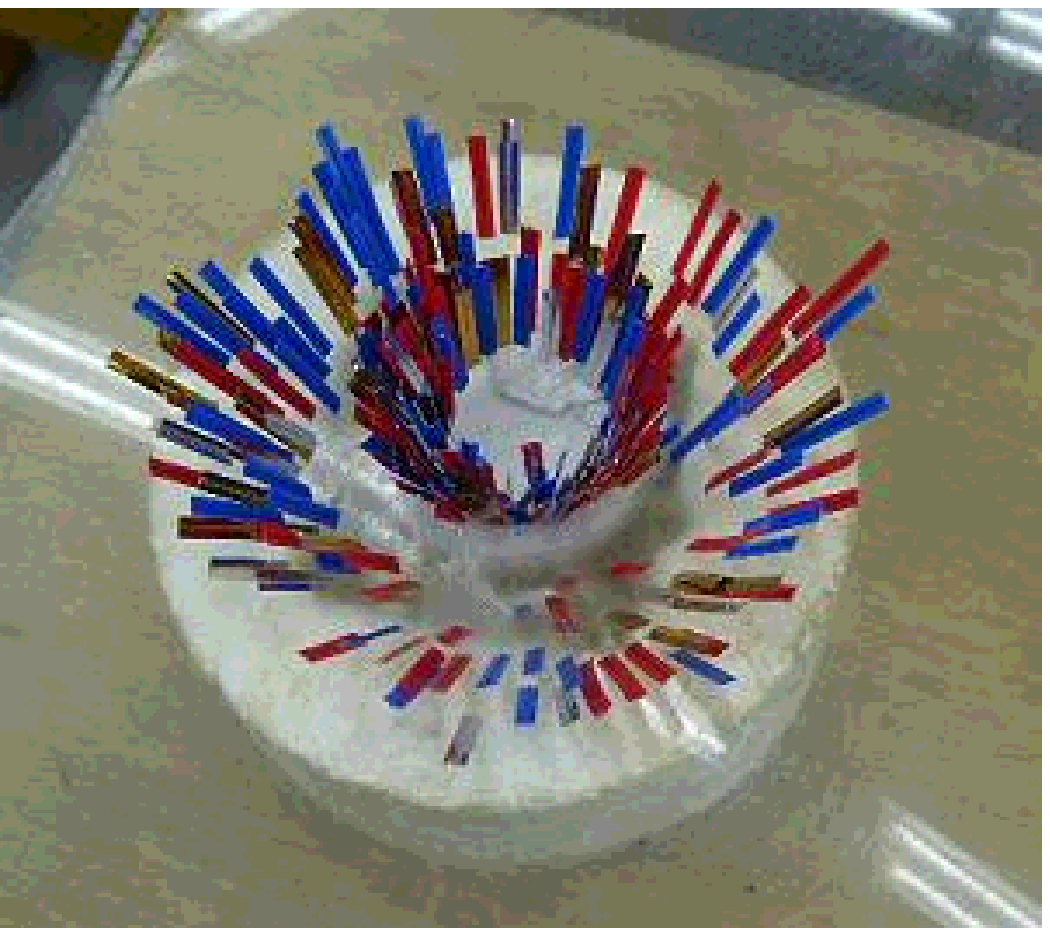}\label{fig:pinning_1b}}
   \end{center}
   \caption{Pinned effect against increasing external magnetic field strength. The magnetic field lines diverging from NdFeB sintered ring magnet are copied inside the HTS. This is on the state of Fig.~\ref{fig:mag_sc2}. The HTS is at a distance of (a)190mm, (b)60mm, from the 100mm cubic NdFeB sintered magnet. The wires are on a transparent plastic board on the top of the HTS. It is shown by the wires that the magnetic field lines penetrate a ring area in the HTS and do not move in it for the change of the external magnetic field strength.
}
    \label{fig:pinning_1}
    
  \begin{center}
    \subfigure[]{\includegraphics[height=40mm]{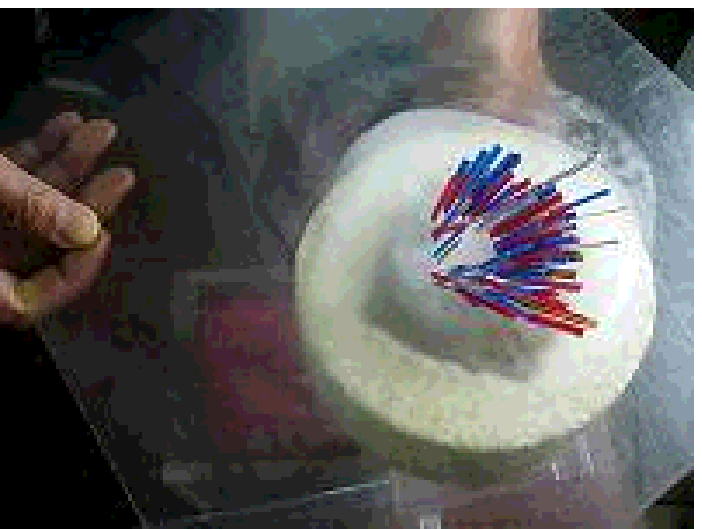}\label{fig:pin_1}}
    \subfigure[]{\includegraphics[height=40mm]{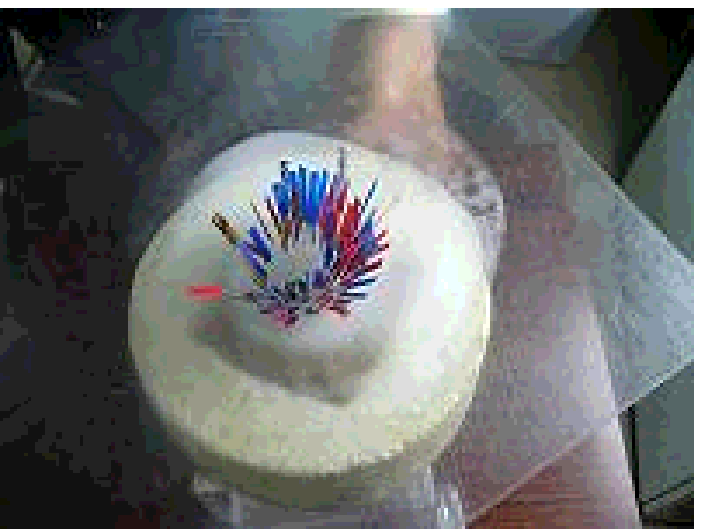}\label{fig:pin_2}}
\subfigure[]{\includegraphics[height=40mm]{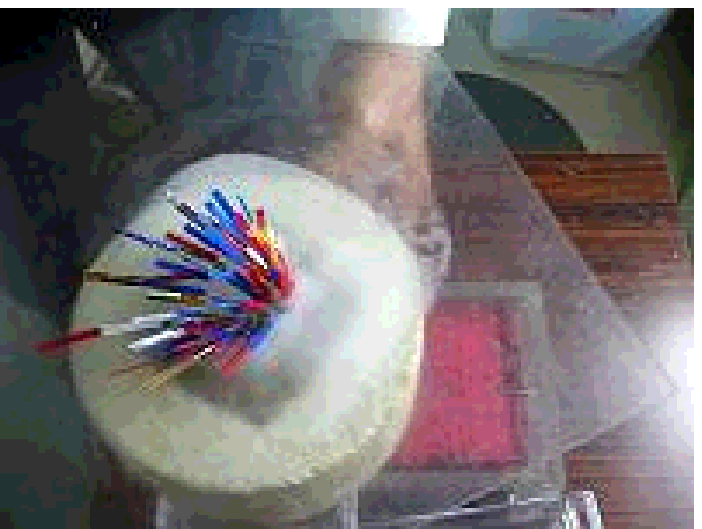}\label{fig:pin_3}}
   \end{center}
   \caption{Pinned effect against tension of external magnetic field. The magnetic field lines diverging from NdFeB sintered ring magnet are copied inside the HTS. This is on the state of Fig.~\ref{fig:mag_sc2}. The HTS is moved in the horizontal direction above the 100mm cubic NdFeB sintered magnet, (a) on the upper right of, (b) just above, (c) on the upper left of, the magnet. The wires are on a transparent plastic board on the top of the HTS. It is shown by the wires that the magnetic field lines penetrate a ring area in the HTS and do not move in it against the tension of the external magnetic field.}
    \label{fig:pinning_2}
\end{figure}

Fig.~\ref{fig:pinning_1} and Fig.~\ref{fig:pinning_2} show the magnetic field lines in the vicinity of the HTS in the state of Fig.~\ref{fig:mag_sc2}. The iron wires are on a transparent plastic board on the top of the HTS. The external magnetic field is caused by the 100mm cubic NdFeB sintered magnet under the HTS.

\begin{figure}
    \begin{center}
    \subfigure[]{\includegraphics[height=40mm]{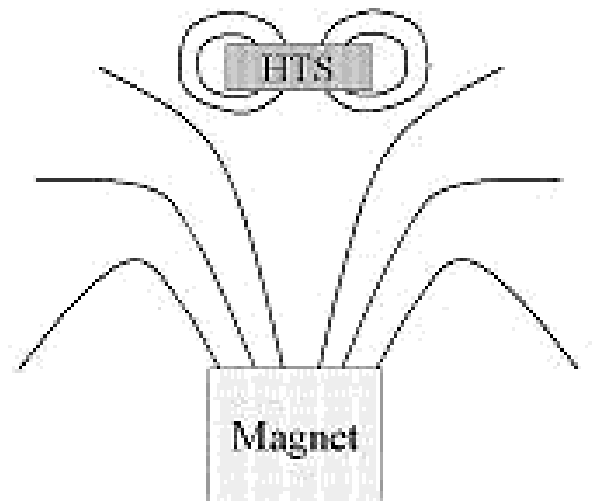}\label{fig:pin_d1}} \subfigure[]{\includegraphics[height=40mm]{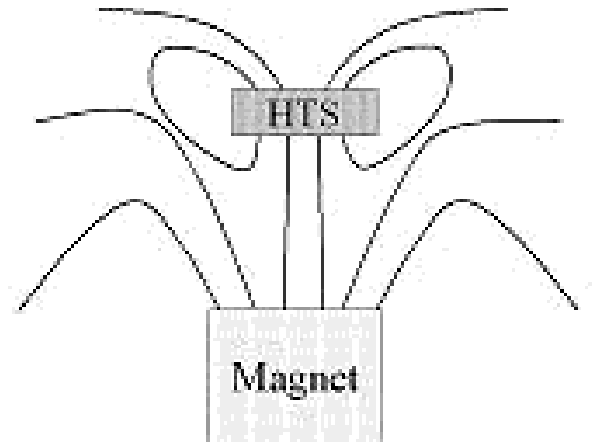}\label{fig:pin_d2}} \subfigure[]{\includegraphics[height=40mm]{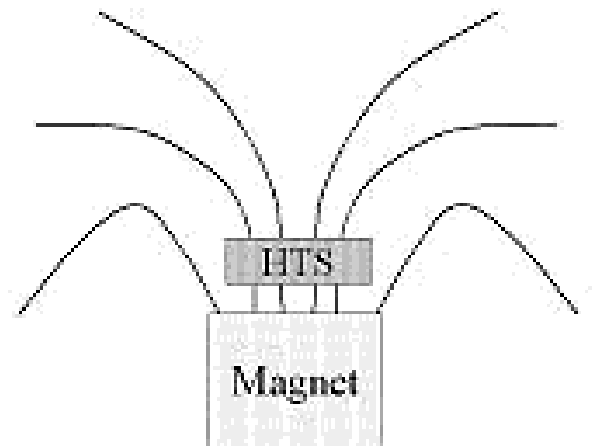}\label{fig:pin_d3}}
   \end{center}
   \caption{Diagram of pinning effect against increasing external magnetic field. If the HTS penetrated by the magnetic field lines is brought close to a magnet, the magnetic field lines penetrating the HTS will begin to be connected to those diverging from the magnet. The magnetic field lines cannot move in a HTS because of the pinning effect. }
    \label{fig:pin_inc_mf}
\end{figure}

Fig .~\ref{fig:pin_inc_mf} is a diagram for Fig.~\ref{fig:pinning_1}. In Fig.~\ref{fig:pinning_1a}, as shown by Fig.~\ref{fig:pin_d1}, the HTS is put at a distance of 190mm from the 100mm cubic NdFeB sintered magnet and the magnetic field lines penetrating the HTS are close near the HTS, because the external magnetic field is negligible for the flux of the penetrating magnetic field lines. In Fig.~\ref{fig:pinning_1b}, as shown by Fig.~\ref{fig:pin_d3}, the HTS is put at a distance of 60mm from the 100mm cubic NdFeB sintered magnet. Some magnetic field lines diverging from the 100mm cubic NdFeB sintered magnet are connected to the magnetic field lines penetrating the HTS and the other magnetic field lines avoid the HTS. It should be emphasized that the magnetic field lines penetrating the HTS are pinned in HTS. 

\begin{figure}
    \begin{center}
    \subfigure[]{\includegraphics[height=40mm]{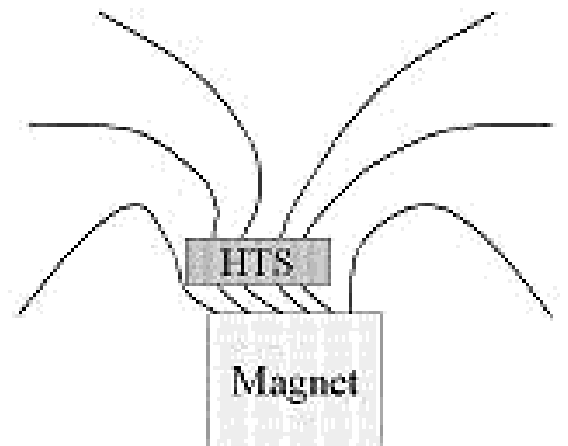}\label{fig:pin_dl}} \subfigure[]{\includegraphics[height=40mm]{pin_d3.eps}\label{fig:pin_dc}} \subfigure[]{\includegraphics[height=40mm]{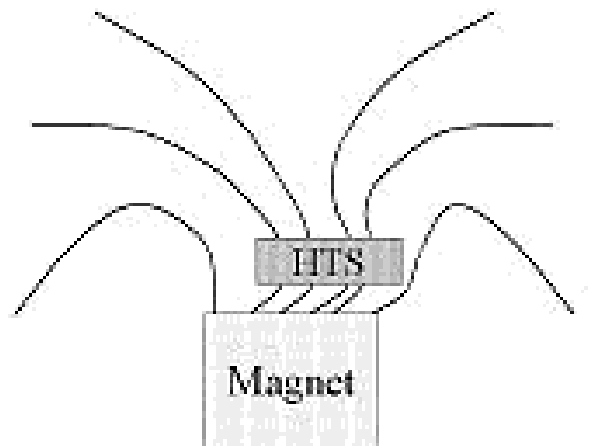}\label{fig:pin_dr}}
   \end{center}
   \caption{Diagram of pinning effect against tension of external magnetic field. If the HTS penetrated by the magnetic field lines is moved in the horizontal direction, the magnetic field lines pinned in the HTS are trawled and made longer. Forces are applied to the magnetic field lines in the HTS to the right (a) or left (c) direction for the tension of the longer magnetic field lines. However, the magnetic field lines cannot move in the HTS because of the pinning effect. 
}
    \label{fig:pin_at_mf}
\end{figure}

Fig .~\ref{fig:pin_at_mf} is a diagram for Fig.~\ref{fig:pinning_2}. If the HTS is moved in the horizontal direction, the magnetic field lines penetrating the HTS are pinned in the HTS, trawled by the movement, and made longer. In Fig .~\ref{fig:pin_1} ( Fig .~\ref{fig:pin_3}), a force to the right (left) direction is applied to the magnetic field lines in the HTS because tension occurs on the longer magnetic field lines as shown by Fig .~\ref{fig:pin_dl} (Fig .~\ref{fig:pin_dr}). It can be seen that the magnetic field lines are pinned against the force. These are the very macroscopic pinned effect.

It can be seen that the magnetic field lines penetrate the HTS, and do not move inside the HTS as the external magnetic field changes.

\newpage 

\begin{figure}
  \centering
    \includegraphics[width=50mm]{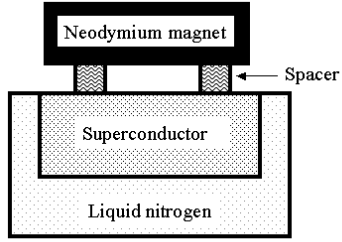}
   \caption{A cooling method to cause Fig.~\ref{fig:mag_sc2}. Before cooling, the magnetic field lines of the magnet penetrate the HTS, and after cooling, they are pinned in the HTS.}
    \label{fig:cooling}
\end{figure}

The HTS had been cooled below the NdFeB sintered ring magnet like Fig.~\ref{fig:cooling}. In other words, the magnetic field lines of the magnet was penetrating at the transition from the normal state to the superconducting state.

If the HTS was a type I superconductor, the magnetic field lines would be excluded from the HTS because the coexisence of the normal state and the superconducting state is forbidden in type I superconductors. In the case of type I\hspace{-.1em}I superconductor, if the external magnetic field strengths is between critical field strengths $H_{c1}$ and $H_{c2}$ ($H_{c1}< H_{c2}$), the quantized flux with normal state core penetrate the superconductor and the vortex lattice is formed by the interactions between the magnetic flux tubes\cite{tonomura}$^,$\cite{Vortex}. This coexisence of the normal state and the superconducting state is called mixed state. If the external magnetic field strengths is more than $H_{c2}$, the superconductor is in the normal state\cite{tinkham}. Though all known HTSs are  type I\hspace{-.1em}I superconductors, the mixed state does not occuer when the magnet get close to the HTS in Fig.~\ref{fig:mag_sc1},  because the magnetic field strengths at the surface of the magnet is less than$ H_{c1}$. If the HTS is cooled like Fig.~\ref{fig:cooling}, the magnetic field lines of the magnet will be penetrate in the HTS as the vortex lattice and will be pinned by the deffects in the HTS crystal. This is just like copying of the magnetic field lines of the magnet to the HTS. Fig.~\ref{fig:pinning_1} shows the "copied" magnetic field lines from the ring magnet. If there was no defect in the HTS crystal, no vortex would be trapped in the HTS. Though it cannot observe the pinned vortex lattice by the method in this paper, Fig.~\ref{fig:pinning_1} and Fig.~\ref{fig:pinning_2} show the pinning effect macroscopically.

\begin{figure}
\includegraphics[height=50mm] {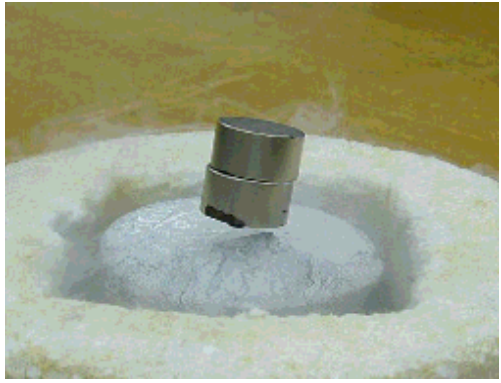}
    \caption{ Rigid levitation of a NdFeB sintered columnar magnet above a superconductor. }
\label{fig:m_s_2a}
\end{figure}

\begin{figure}
  \begin{center}
    \subfigure[]{\includegraphics[height=40mm]{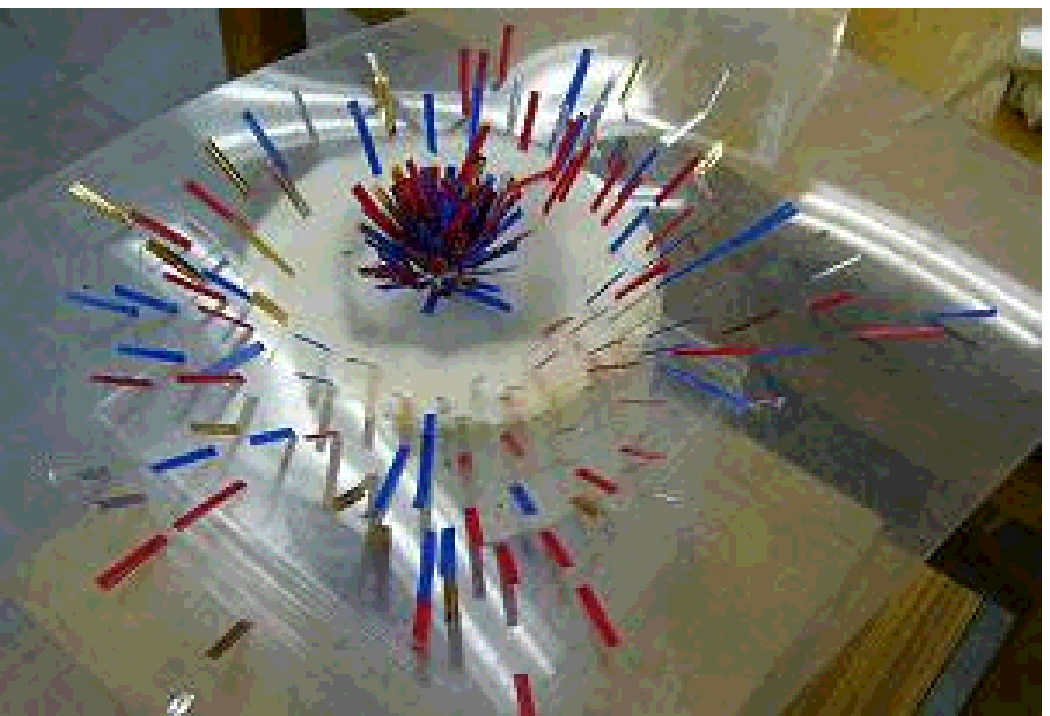}\label{fig:pinning_2b}}
    \subfigure[]{\includegraphics[height=40mm]{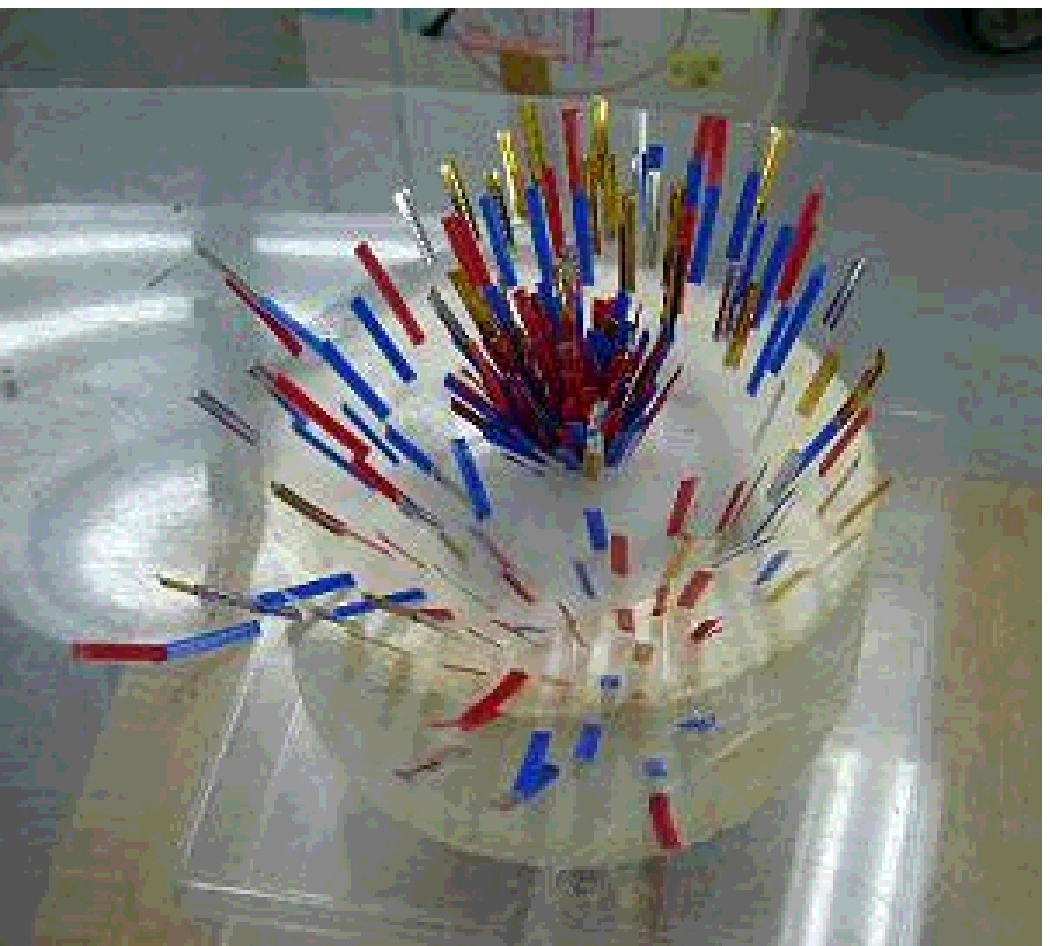}\label{fig:pinning_2c}}
    \subfigure[]{\includegraphics[height=40mm]{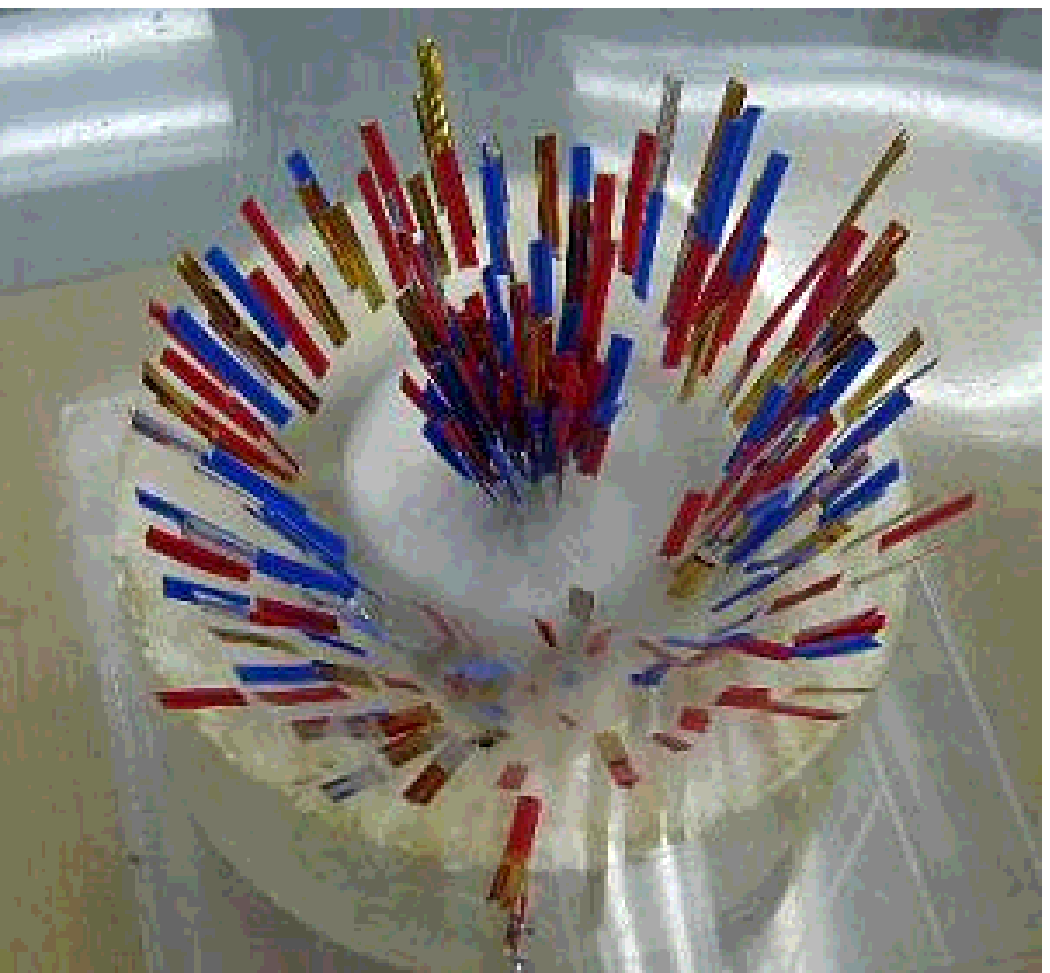}\label{fig:pinning_2d}}
   \end{center}
   \caption{Pinned magnetic field lines of a NdFeB sintered columnar magnet. The HTS is at a distance of  (a)190mm, (b)150mm, (c)100mm, from  the 100mm cubic NdFeB sintered magnet. The wires are on a transparent plastic board on the top of the HTS. It is shwon that the magnetic field lines penetrate a circle area in the HTS and do not move in it for the change of the external magnetic field.} 
\label{fig:pinning_3}
\end{figure}

Fig.~\ref{fig:m_s_2a} is the case that the HTS had been cooled below the NdFeB sintered columnar magnet like Fig.~\ref{fig:cooling} and a rigid levitation of a NdFeB sintered columnar magnet above a superconductor occurred. In Fig.~\ref{fig:pinning_2b}, Fig.~\ref{fig:pinning_2c}, and Fig.~\ref{fig:pinning_2d}, the HTS is at a distance of 190mm, 150mm, and100mm respectively from the 100mm cubic NdFeB sintered magnet. The flux of the magnetic field copied in the HTS are comparable to that of the external magnetic field. It may be seen clearer than Fig.~\ref{fig:pinning_1} that the magnetic field lines do not move as the external magnetic field changes. 

Movies show much more clearly than these figures\cite{www}. These represent the pinning effect qualitatively.

\subsection{Superconduction state breaking }
\begin{figure}[]
  \centering
    \subfigure[]{\includegraphics[height=40mm]{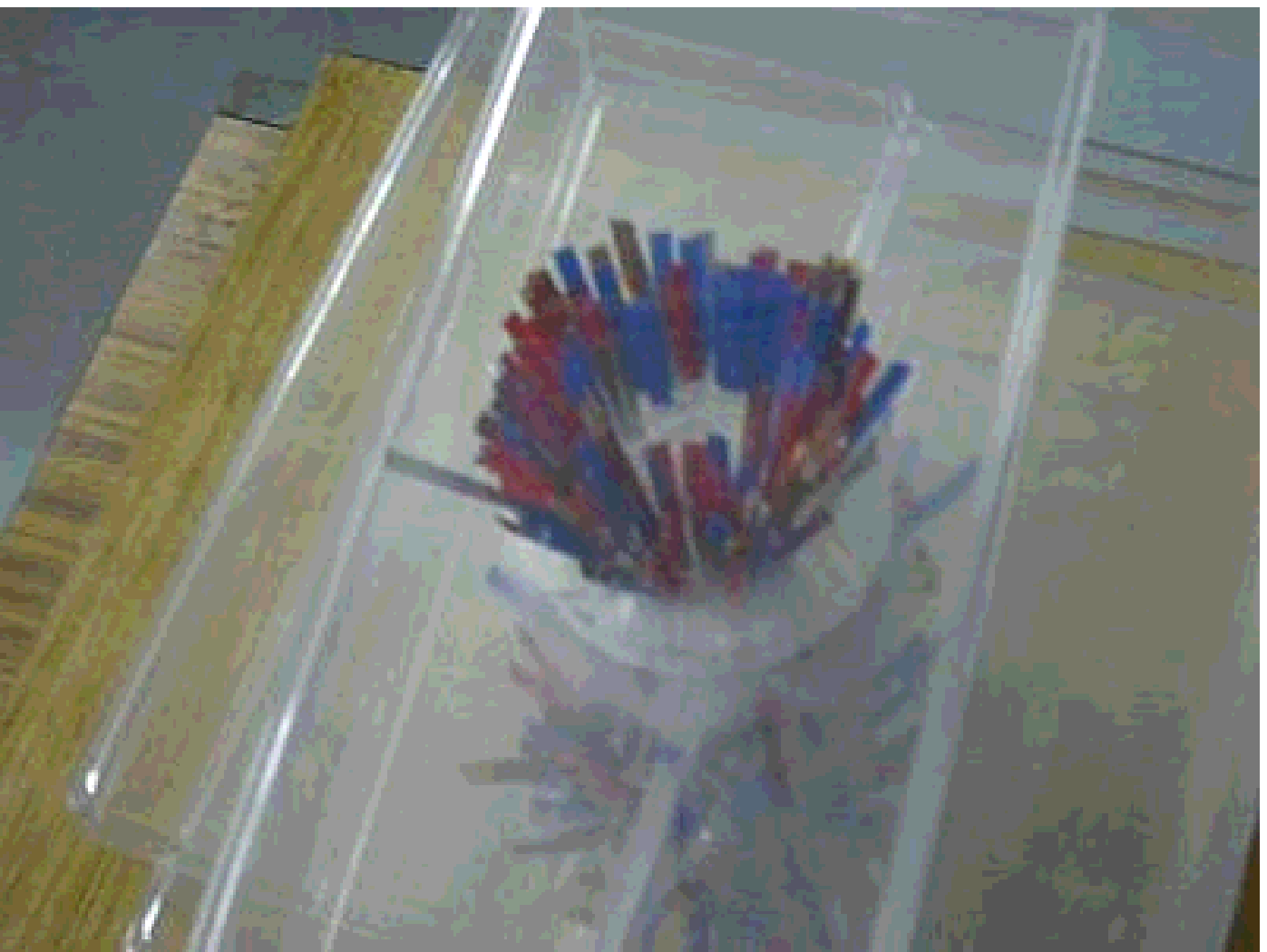}\label{fig:warming_a}}
    \subfigure[]{\includegraphics[height=40mm]{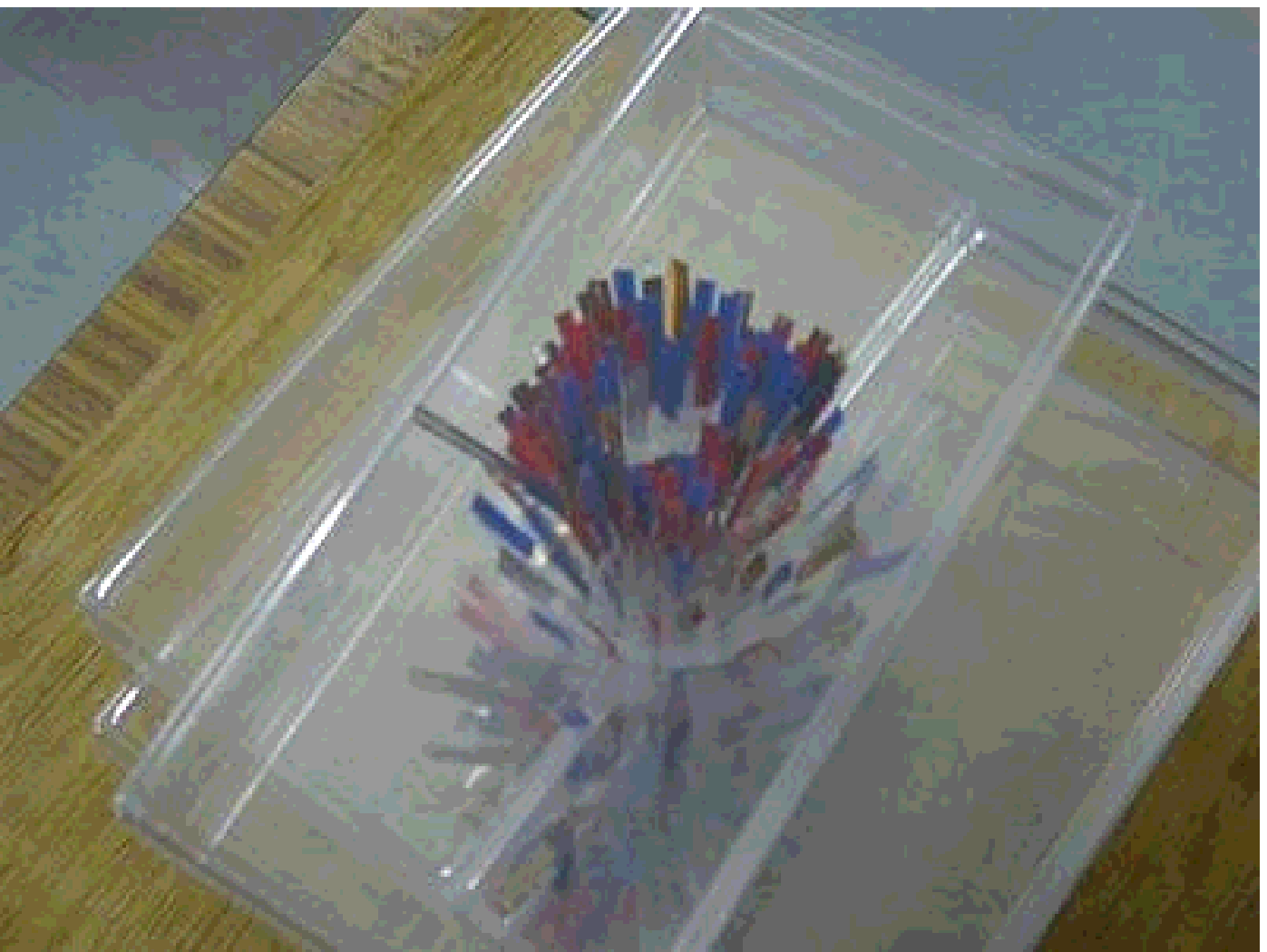}\label{fig:warming_b}}
    \subfigure[]{\includegraphics[height=40mm]{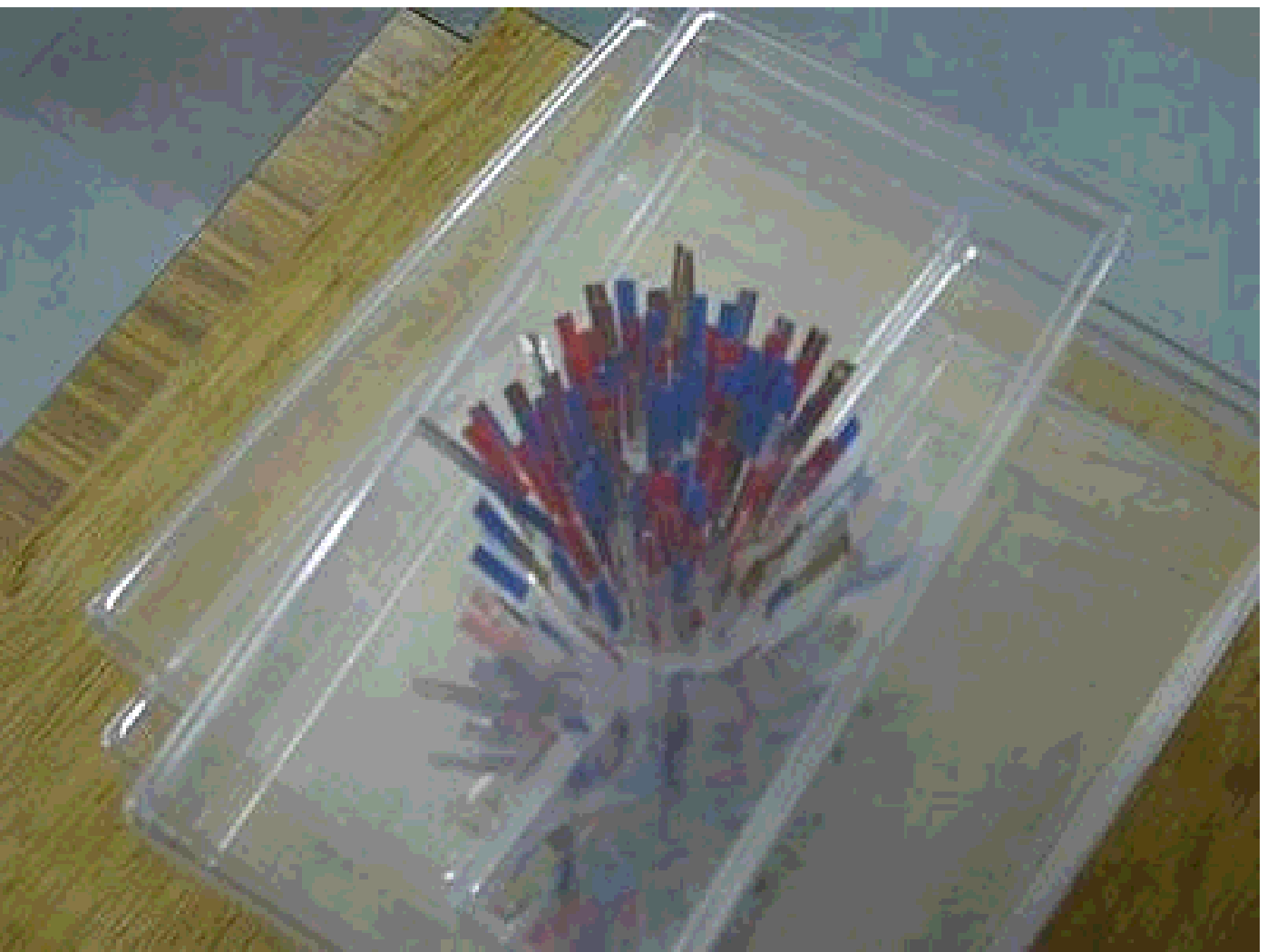}\label{fig:warming_c}}
   \caption{Superconduction state breaking.
The wires show the change of the area penetrated by the magnetic field lines in the HTS.
The HTS with pinned magnetic field lines of the NdFeB sintered ring magnet is, (a) put below the wires, and above the 100mm cubic NdFeB sintered magnet in the room temperature,
(b) warming in the room temperature, and (c) at the temperature of normal conduction state.}
    \label{fig:warming_sc}
\end{figure}

Fig.~\ref{fig:warming_sc} represents a breaking superconduction state when the HTS in Fig.~\ref{fig:pinning_1} is warmed in the room temperature. Fig.~\ref{fig:warming_a} show the magnetic field lines "copied" from the ring magnet on the top of the HTS. The area where the magnetic field lines do not penetrate is in the superconduction state. The area where the magnetic field lines penetrate is in the mixed state, that is, the vortex lattice is formed in this area. In Fig.~\ref{fig:warming_b}, on the one hand, the superconduction state area seems to decrease, on the other, the mixed state area seems to increase. This is why the critical field strengths $H_{c1}$ and $H_{c2}$ decrease with the HTS temperature. To be exact, expect  the area copied the magnetic field lines from the ring magnet, the area is the superconduction state, the mixed state, and the normal state when $ H_{E}<H_{c1}$, $H_{c1}<H_{E}< H_{c2}$, $H_{c2}<H_{E}$ respectively, where  $H_{E}$ is the external magnetic field strength. In Fig.~\ref{fig:warming_c}, the magnetic field lines penetrate all over the HTS. If $H_{c2}<H_{E}$, then the whole HTS is in the normal conduction state. 

A movie shows this phenomenon of the superconduction state breaking much more vividly than these figures\cite{www}. The phenomena are the transition from the superconduction state to the mixed state and from the mixed state to the normal state. If the pinning effect depends on the temperature, this may be seen from the observation of this superconduction state breaking.      .

\section{conclusion}
It was shown the very simple method to observe the magnetic field lines for the Meissner effect as well as for the pinning effect. The author prepared some movies on a webpage\cite{www}. The movies have much more plentiful supply of information than the photographs in this article. Even the beginners can clearly observe the behavior of the magnetic field lines related superconductivity. This observation is effective for the beginners. In fact, many visitors in Osaka Science Museum enjoy not only the unexpected phenomena of the interaction between a superconductor and a magnet such as Fig.~\ref{fig:mag_sc1} and Fig.~\ref{fig:mag_sc2}, but also considering the interaction based on the magnetic field lines. Thus the beginners enjoy the essence of superconductivity. This observation must be useful at any introduction of superconductivity because the method is very simple and easy to set up and to use, and the phenomena is enough large and clear to demonstrate with the naked eye.

\begin{acknowledgments}
The author would like to express his gratitude to Nippon Steel Corporation and Dr. Hidekazu Teshima (Advanced Technology Research Laboratories, Nippon Steel Corporation). Nippon Steel Corporation sympathized with the author's design of the demonstration and supplied the GdBaCuO bulk HTS. Dr. H. Teshima coped with the difficulties for the author. 
\end{acknowledgments}

\end{document}